\documentclass[trackchanges,twocolumn,tighten
]{aastex701}
\usepackage{amsmath,multirow}
\hypersetup{linkcolor=red,citecolor=green,filecolor=cyan,urlcolor=magenta}

\begin{document}

\title{Stationary Power-Law Solutions of Weak Kinetic-Alfv\'{e}nic Turbulence}

\author[orcid=0000-0002-0512-6273]{Kexun Shen}
\affiliation{Institute for Fusion Theory and Simulation, School of Physics, Zhejiang University, Hangzhou 310027, China}
\email{kxshen10@zju.edu.cn}  

\author[orcid=0009-0006-1426-0673]{Zhiwen Cheng}
\affiliation{Institute for Fusion Theory and Simulation, School of Physics, Zhejiang University, Hangzhou 310027, China}
\email{12245071@zju.edu.cn}  

\author[orcid=/0000-0002-7548-8819]{Zhiyong Qiu}
\affiliation{CAS Key Laboratory of Frontier Physics in Controlled Nuclear Fusion and Institute of Plasma Physics, Chinese Academy of Sciences, Hefei 230031, China}
\affiliation{Center for Nonlinear Plasma Science, ENEA C.R. Frascati, Frascati 00044, Italy}
\email[show]{zqiu@ipp.ac.cn}  

\correspondingauthor{Zhiyong Qiu}


\begin{abstract}
The wave-kinetic description of weak kinetic-Alfv\'{e}nic turbulence based on the gyrokinetic theoretical framework is proposed. The wave kinetic equation describing kinetic Alfv\'{e}n wave spectral cascading via resonant three-wave interactions is derived, and the stationary spectra are analytically obtained using the Zakharov transformation in both the long-wavelength limit and the short-wavelength limit, for both counter-propagating and co-propagating cases. The cascade directions of stationary solutions are identified and their existence is further verified by numerical solution of the wave kinetic equation. A brief discussion on the relevance of such predictions to the solar wind turbulence and helical kinetic-Alfv\'{e}nic turbulence is presented.
\end{abstract}


\section{Introduction}\label{sec:intro}

Plasma turbulence characterized by low-frequency shear-Alfv\'{e}nic oscillations is ubiquitous in space, astrophysical and laboratory plasma environments, such as the solar wind\citep{RBrunoLRSP2013}, the solar corona\citep{SRCranmerARAA2019}, interstellar media\citep{BGElmegreenARAA2004}, and magnetically confined fusion devices\citep{LChenRMP2016}. 
Among them, the solar wind has been found to be highly turbulent with power-law spectra spanning a vast range of spatial scales, e.g.,  $E_k\propto k_{\perp}^{x}$, with $E_k$ being the turbulent energy density, $k_{\perp}$ being the wavenumber perpendicular to the mean magnetic field, and $x$ being the power-law index. The Alfv\'{e}nic fluctuations are characterized by frequencies below the ion cyclotron frequency and the strong anisotropy in correlation length with respect to the mean magnetic field\citep{HAlfvenNature1942}. At magnetohydrodynamic (MHD) scales where the dissipation of Alfv\'{e}n waves (AWs) is negligible, turbulent energy are transferred to the ion gyroscale where AWs become dispersive and can be damped. 
The turbulent cascade at kinetic scales attributed to kinetic Alfv\'{e}n waves (KAWs)\citep{AHasegawaPoF1976,LChenRMPP2021} has brought up new issues on the multi-scale spectral feature of the solar wind, the intermittent structure formation\citep{SBoldyrevApJL2012,MZhouPNAS2023}, turbulent heating processes\citep{CHKChenNatCommun2019,RMeyrandPNAS2019,TABowenPRL2022,JSquireNatAstron2022}, and weak turbulence theories\citep{RMeyrandPRL2016,VDavidPRL2024}. 

For a strongly imbalanced solar wind turbulence, the formation of a spectral transition range with a power-law index of around $-4$  and the plasma heating at the ion gyroscale are found to be related to the conservation of energy and helicity of KAW turbulence, as suggested by recent Parker Solar Probe (PSP) observations\citep{TABowenNatAstron2024}. Here, ``imbalance" corresponds to the case when Alfv\'{e}nic fluctuations outward and inward-propagating from the Sun differ, and could be quantified by the helicity invariant. At kinetic scales, the helicity is subject to an inverse cascade towards large scales\citep{JChoPRL2011,GMiloshevichJPP2021}, which prevents the KAW turbulence from cascading to sub-ion scales and results in the generation of ion cyclotron waves that heat background plasmas\citep{RMeyrandJPP2021,JSquireNatAstron2022,TABowenNatAstron2024}. 
On the other side, regardless of dissipations, nonlinear interactions among co-propagating KAWs may also lead to such a steep spectrum near the ion gyroscale\citep{TABowenPRL2020,TPassotJPP2022,TPassotFPP2024}, which indicates a correlation between the steepness of the slope and the degree of imbalance\citep{SYHuangApJL2021}. On the theoretical aspect, the roles of the helicity and the imbalance of solar wind have been intensely studied in the strong turbulence regime using phenomenological arguments mostly, while remain to be explored in the weak turbulence regime. 

Over the past two decades, investigations on the kinetic turbulent cascade of weak Alfv\'{e}nic turbulence have mostly concentrated on the direct cascade towards small scales for the balanced case\citep{SGaltierJPP2000,SBoldyrevPRL2009,SGaltierJPP2015,VDavidPRL2024}, which is seen as the result of successive collisions between counter-propagating wave-packets. Here, ``collision" indicates the interaction among different waves through nonlinear mode-mode coupling processes, when one considers the waves as quasi-particles. Based on standard perturbative treatments and the multiple-time asymptotic closure, the kinetic equation for waves can be obtained to describe the nonlinear dynamics of randomly distributed dispersive waves and the long-time statistical behavior of physical quantities like the energy, the momentum and the wave-action\citep{ACNewellPhysicaD2001,SNazarenkoCP2015}. 
Several fluid models related to the KAW turbulence have been applied to weak turbulence analyses. For example, the nonlinear evolution of weakly interacting shear-Alfv\'{e}n wave-packets have been analyzed using the reduced magnetohydrodynamics (RMHD) in isotropic\citep{RHKraichnanPoF1965} and anisotropic cases\citep{CSNgApJ1996,SGaltierApJ2002}. At sub-ion scales, electron reduced magnetohydrodynamics (ERMHD) contains KAWs with Boltzmannian ions and massless isothermal fluid electrons\citep{AASchekochihinApJS2009,SGaltierJPP2015}. 
Although only counter-propagating AWs can interact with each other in the ideal MHD uniform plasma limit, co-propagating KAWs may also lead to effective nonlinear energy transfer due to the finite-ion-Larmor-radius effect that breaks the ``pure Alfv\'enic state" \citep{WMElsasserRMP1956,LChenPoP2013}. Within the parametric decay analysis\citep{YMVoitenkoJPP1998,KShenPoP2024} which reveals the fundamental property of three-wave interaction, the dual-cascade behavior is identified for co-propagating KAW triplets, in analogy to the two-dimensional hydrodynamic turbulence\citep{RHKraichnanPoF1967} and the drift wave turbulence in magnetized plasmas described by the Charney-Hasegawa-Mima (CHM) equation\citep{AHasegawaPoF1978,CConnaughtonPR2015}. Both the two-dimensional Navier-Stokes (NS) equation and the CHM equation possess two positive quadratic invariants, i.e. the energy and the enstrophy. The inverse energy cascade process and selective dissipations due to the conservation of enstrophy result in the development of self-organized states in these quasi-two-dimensional turbulent systems\citep{RHKraichnanRPP1980,AHasegawaAP1985,GBoffettaARFM2012,AAlexakisPR2018,AAlexakisRMPP2023}. 
Due to the existence of conserved energy and helicity, the theory of weak kinetic-Alfv\'{e}nic turbulence may shed light on the realistic cross-scale evolution of solar wind turbulence by giving mathematical explanations to cascade behaviors, exact stationary power-law spectra, non-stationary self-similar solutions, etc\citep{VEZakharov1992book,SNazarenko2011book,ACNewellARFM2011}, and this constitutes the motivation of the present work. 

In the present study, we adopt the gyrokinetic formulation of KAW turbulence in order to fully retain the effect of ion Larmor radius, crucial for the proper description of KAWs linear properties and nonlinear cascading. In Section \ref{sec:model-equation}, a general nonlinear mode equation for low-$\beta$ KAW turbulence is given. In Section \ref{sec:methodology}, we provide the general methodology for weak KAW turbulence including the description of the wave kinetic equation, the derivation of the stationary spectra of axisymmetric systems, and the identification of cascade direction of each stationary solution. In Section \ref{sec:results} we report our main analytical results with supplementary numerical calculations. Finally we have a discussion and the conclusion of the study in Section \ref{sec:conclusion-discussion}. 

\section{Gyrokinetic Model Equations}\label{sec:model-equation}

The dynamics of low-frequency Alfv\'{e}nic fluctuations around ion-gyroscale can be rigorously described from a reduction of gyrokinetics given here.  Within the scope of weak turbulence theory, the governing  equation describing the nonlinear interactions among KAWs are formulated in Refs. \cite{YMVoitenkoJPP1998,KShenPoP2024}. Here, for the self-containess of the paper and the convenience of readers, we will briefly give the derivation of the nonlinear mode equation that will be used in the following sections for the KAW spectral cascading. Following \cite{KShenPoP2024}, we consider a  uniform plasma near Maxwellian equilibria for different particle species $s=i,e$, in which case the particle distribution function can be decomposed into a Maxwellian distribution $F_{Ms}$ and a perturbed component $\delta f_s$. The plasma is embedded in a uniform mean magnetic field $\mathbf{B}_0=B_0\hat{\mathbf{b}}_0$ with $\hat{\mathbf{b}}_0=\hat{\mathbf{e}}_{\parallel}$ being the unit vector along the  mean magnetic field. Furthermore, Coulomb collisions are also neglected. KAWs are characterized by low-frequency $\omega_k\ll\Omega_{i}$ electromagnetic perturbations which are highly anisotropic with respect to the mean field $k_{\parallel}\ll |\mathbf{k}_{\perp}|$, where $\Omega_{s}$ is the gyrofrequency and $(\omega_k,\mathbf{k})$ denotes the four-wavevector of a discrete mode in the Fourier decomposition. Assuming low-$\beta$ limit with $\beta$ being the ratio between thermal and magnetic pressures, the high-frequency compressive Alfv\'{e}nic fluctuations are omitted in our model and the perturbed fields can be expressed in terms of the scalar potential $\delta\phi$ and the parallel magnetic vector potential $\delta A_{\parallel}$ \footnote{It is noteworthy that, the low-$\beta$ limit may sometimes fail for space/solar plasmas of interest. We, however, still assume the low-$\beta$ limit for the simplicity of the analysis.}. Given the formal gyrokinetic ordering\citep{AJBrizardRMP2007}, the perturbed distribution function can be written as 
\begin{equation}
    \delta f_s=-\frac{q_s\delta\phi}{T_{s}}F_{Ms}+e^{-\mathbf{\rho}_s\cdot\nabla}\delta g_s,
\end{equation}
consisting of the adiabatic particle response and the gyroangle-independent non-adiabatic response $\delta g_s$, with $\exp (-\mathbf{\rho}_s\cdot\nabla)$ denoting the generator of coordinate transformation from guiding-center phase space to particle space, $q_s=\pm e$ and $T_s$ being the charge and mean temperature for ions and electrons. The non-adiabatic response satisfies the collisionless nonlinear gyrokinetic equation\citep{EAFriemanPoF1982,LChenJGR1991}: 
\begin{equation}\label{eqn:NLGKE}
    (\partial_t+v_{\parallel}\nabla_{\parallel})\delta g_s+\frac{c}{B_0}[\langle\delta L_g\rangle_{\alpha},\delta g_s]=\frac{q_sF_{Ms}}{T_{s}}\partial_t\langle\delta L_g\rangle_{\alpha},
\end{equation}
where $\langle \delta L_g\rangle_{\alpha}\equiv\langle \exp (\mathbf{\rho}_s\cdot\nabla)\delta L\rangle_{\alpha}$, $\delta L\equiv \delta\phi -v_{\parallel}\delta A_{\parallel}/c$ is the effective potential, $\langle ... \rangle_{\alpha}$ denotes gyro-averaging, $\langle\exp (\mathbf{\rho}_s\cdot\nabla)\rangle_{\alpha}=J_{0s}(k_{\perp}\rho_s)$ with $k_{\perp}^2=-\Delta_{\perp}$ being an operator, and $J_0$ being the zeroth-order Bessel function accounting for finite Larmor radius effects. The Poisson bracket is defined as $[f,g]=\hat{\mathbf{b}}_0\cdot\nabla_{\perp} f\times\nabla_{\perp} g$ representing the nonlinear mode-coupling. The appended field equations are 
the parallel Ampere's law 
\begin{equation}\label{eqn:||Ampere}
    \delta J_{\parallel}=-\frac{c}{4\pi}\Delta_{\perp}\delta A_{\parallel}=\sum_sq_s\langle v_{\parallel}J_{0s}\delta g_s\rangle_v,
\end{equation}
where $\langle ...\rangle_v$ is the velocity integration over $(v_{\parallel},v_{\perp})$, and the quasi-neutrality constraint 
\begin{equation}\label{eqn:QNcond}
    \sum_sq_s\delta n_s=\sum_sq_s\left(-\frac{q_s\delta\phi}{T_{s}}n_{0}+\langle J_{0s}\delta g_s\rangle_v\right)=0,
\end{equation}
which gives 
\begin{equation}
    \delta n_e=\delta n_i\simeq \frac{n_0e}{T_i}(\Gamma_0-1)\delta\phi,
\end{equation}
after the substitution of the linear non-adiabatic ion response $\delta g_i\simeq (eF_{Mi}/T_i)J_{0i}\delta\phi$ in the limit of $v_{ti}/v_A\sim\sqrt{\beta_i}\ll 1$. Here, $n_0$ is the plasma density, and $\Gamma_0=I_0(b_0)\exp (-b_0)$ represents the full finite-ion-Larmor-radius (FILR) effect with $I_0(b_0)$ the zeroth-order modified Bessel function, $b_0=-\rho_i^2\Delta_{\perp}$, $\rho_i=v_{ti}/\Omega_i$ is the ion Larmor radius, $v_{ti}=\sqrt{T_{i}/m_i}$ is the ion thermal velocity, and $v_A=B_0/\sqrt{4\pi n_0m_i}$ is the Alfv\'en velocity. The generalized parallel Ohm's law can be obtained from the electron gyrokinetic equation in the $\beta_e\gg m_e/m_i$ limit, by neglecting the electron inertia and substituting in the particle response $\langle\delta g_e\rangle_v=\delta n_e-(n_0e/T_e)\delta\phi$:
\begin{equation}\label{eqn:||Ohm}
    c^{-1}\partial_t\delta A_{\parallel}+\nabla_{\parallel}\sigma_0\delta\phi=B_{0}^{-1}[\delta A_{\parallel},\sigma_0\delta\phi],
\end{equation}
with $\sigma_0\equiv 1+\tau (1-\Gamma_0)$ an operator and $\tau\equiv T_e/T_i$.
Multiplying Equation (\ref{eqn:NLGKE}) by $q_sJ_{0s}$, taking the velocity-space integral and making use of Equation (\ref{eqn:||Ampere}), one has the generalized nonlinear gyrokinetic vorticity equation\citep{LChenNF2001,LChenRMPP2021}:
\begin{equation}\label{eqn:NLGKV}
    \begin{aligned}
    & (\Gamma_0-1)\partial_t\delta\phi+\frac{\rho_i^2v_A^2}{c}\nabla_{\parallel}\Delta_{\perp}\delta A_{\parallel}=\\ & -\frac{c}{B_0}[\delta\phi,\Gamma_0\delta\phi]+\frac{\rho_i^2v_A^2}{cB_0}[\delta A_{\parallel},\Delta_{\perp}\delta A_{\parallel}].
    \end{aligned}
\end{equation}
The two terms on the left hand side of Equation (\ref{eqn:NLGKV}) correspond to the generalized inertia and field line bending terms, while the two terms on the right hand side are the formally nonlinear terms, i.e., the generalized gyrokinetic Reynolds stress and Maxwell stress, respectively.
The finite Larmor radius effect due to electrons is neglected assuming $k_{\perp}\rho_e\sim k_{\perp}\rho_i(m_e/m_i)^{1/2}\ll 1$ with $k_{\perp}\rho_i\sim\mathcal{O}(1)$.  

The model given here, consisting of the generalized nonlinear gyrokinetic vorticity equation and the generalized parallel Ohm's law, is derived from the nonlinear gyrokinetic theory framework including the quasi-neutrality condition and the parallel Ampere's law in the limit of $1\gg\beta_i\sim\beta_e\gg m_e/m_i$, and is equivalent to the nonlinear mode equation derived in \cite{KShenPoP2024}. It serves as the minimal model for nonlinear dynamics and multiscale turbulent evolution of low-frequency Alfv\'{e}nic oscillations retaining the complete FILR effect. 
Combining Equations (\ref{eqn:||Ohm}) and (\ref{eqn:NLGKV}), one then obtains the desired nonlinear mode equation describing the nonlinear spectral evolution of KAW turbulence in its Fourier form:
\begin{equation}\label{eqn:nl_mode_eqn}
    b_k\epsilon_{Ak}\delta\phi_k=\int\frac{i\Lambda_{12}^k}{2\omega_k}\beta_{12}^k\delta\phi_1\delta\phi_2\delta_{12}^kd\mathbf{k}_1d\mathbf{k}_2,
\end{equation}
with $\Lambda_{12}^k\equiv\hat{\mathbf{b}}_0\cdot\mathbf{k}_{1\perp}\times\mathbf{k}_{2\perp}$,  
\begin{equation}
    \epsilon_{Ak}=\frac{1-\Gamma_k}{b_k}-\frac{k_{\parallel}^2v_A^2}{\omega_k^2}\sigma_k,
    \end{equation}
being the linear dielectric constant of KAW from which the KAW linear dispersion relation can be obtained,
\begin{equation}
    \begin{aligned}
        \beta_{12}^k= & v_A^2\sigma_1\sigma_2\left(\frac{k_{2\parallel}}{\omega_2}-\frac{k_{1\parallel}}{\omega_1}\right)\\ & \times\left(b_k\frac{k_{\parallel}}{\omega_k}+b_1\frac{k_{1\parallel}}{\omega_1}+b_2\frac{k_{2\parallel}}{\omega_2}\right),
    \end{aligned}
\end{equation}
being the nonlinear coupling coefficient and $\delta_{12}^k\equiv\delta (\mathbf{k}-\mathbf{k}_1-\mathbf{k}_2)$ indicating the wavevector resonance condition of three-wave interaction $\mathbf{k}-\mathbf{k}_1-\mathbf{k}_2=0$. We note that, Equation (\ref{eqn:nl_mode_eqn}) was originally derived in Refs. \cite{YMVoitenkoJPP1998,KShenPoP2024}, and has an interesting correspondence to the famous CHM equation describing turbulent cascading of drift waves \citep{AHasegawaPoF1978}. Equation (\ref{eqn:nl_mode_eqn}) can be applied to investigate both the coherent mode-coupling processes like the parametric decay process\citep{KShenPoP2024}, and turbulence studies including weak and strong turbulent regimes. In the rest of the paper we focus on the weakly turbulent behavior of KAWs considering resonant three-wave interactions. 

Equations (\ref{eqn:||Ohm}) and (\ref{eqn:NLGKV}), and thus, Equation (\ref{eqn:nl_mode_eqn}), conserve two quadratic invariants, i.e. the generalized free-energy\citep{AASchekochihinApJS2009}
\begin{equation}\label{eqn:generalized_free_energy}
    E=\int d\mathbf{r}\left[\frac{n_0e^2}{T_{i}}\frac{\sigma_0(1-\Gamma_0)|\delta\phi|^2}{2}+\frac{|\nabla_{\perp}\delta A_{\parallel}|^2}{8\pi}\right],
\end{equation}
which is composed of equi-partitioned kinetic energy and magnetic energy, and the generalized helicity\citep{TPassotPoP2018,RMeyrandJPP2021}
\begin{equation}\label{eqn:generalized_helicity}
    P=\int d\mathbf{r}\delta A_{\parallel}\frac{1-\Gamma_0}{\rho_i^2}\delta\phi,
\end{equation}
which can be reduced to the cross-helicity of SAW turbulence in the MHD limit.  The conservation of $E$ and $P$ determines the dynamic  evolution of the system, and the impact will be analyzed below.

\section{Weak Turbulence Spectrum Methodology}\label{sec:methodology}

Based on the general nonlinear mode equation (\ref{eqn:nl_mode_eqn}) describing the KAW turbulence energy transfer and cascading behaviors, the methodology of deriving the stationary spectra of weak KAW turbulence is proposed here. We firstly deliver a statistical description of the weak kinetic-Alfv\'{e}nic turbulence in Section \ref{sec:WKE} adopting the quasi-particle picture of plasma waves. On the basis of the kinetic equation for waves, the stationary power-law spectra applicable for axisymmetric anisotropic media\citep{EAKuznetsovJETP1972,VEZakharov1992book} is derived in Section \ref{sec:KZ-spectra} and their cascade directions are discussed correspondingly in Section \ref{sec:Cascade-Direction}. The application to the analysis of KAW stationary spectra in various parameter regimes will be presented in Section \ref{sec:results}.

\subsection{Wave Kinetic Equation}\label{sec:WKE}

\begin{figure*}[htbp!]
    \centering
    \begin{minipage}{.49\textwidth}
        \centering 
        \includegraphics[width=0.99\linewidth]{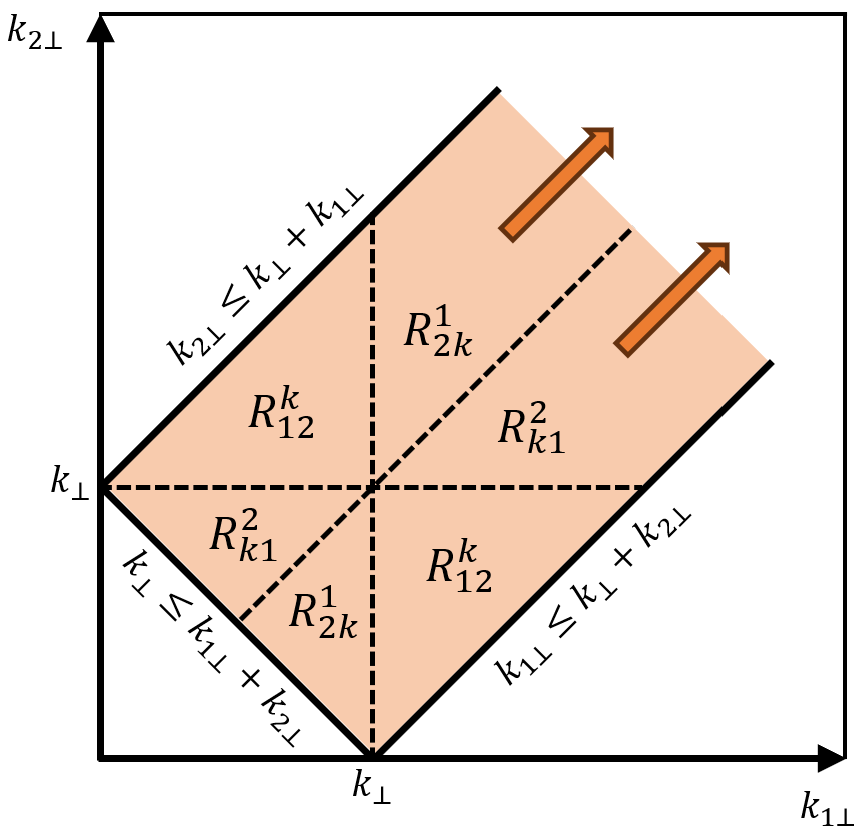}
        \caption{\label{fig:match_cond_co-prop} Integration region of each integrand for the co-propagating case $(s_k=s_1=s_2)$. The resonant three-wave interaction among co-propagating KAWs exhibits a dual-cascading character.}
    \end{minipage}
    \begin{minipage}{.49\textwidth}
        \centering 
        \includegraphics[width=0.99\linewidth]{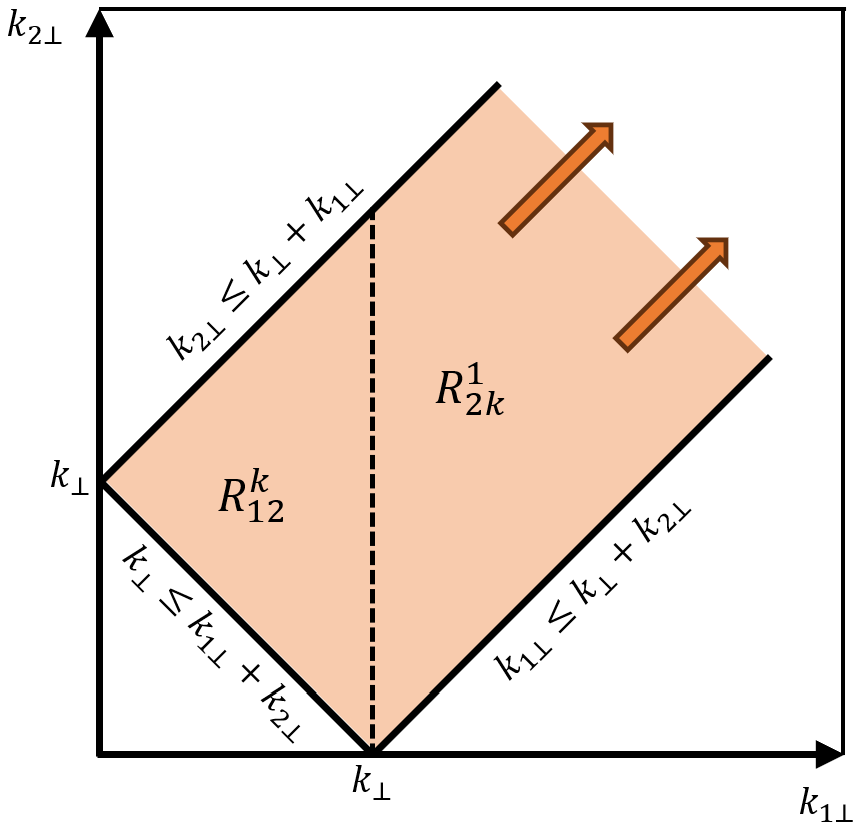}
        \caption{\label{fig:match_cond_counter-prop} Integration region of each integrand for the counter-propagating case $(s_k=s_1=-s_2)$. The resonance conditions indicate that $\mathcal{R}_{12}^k$ and $\mathcal{R}_{2k}^1$ are defined on two intervals $k_{\perp}>k_{1\perp}$ and $k_{\perp}<k_{1\perp}$, respectively.}
    \end{minipage}
\end{figure*}

Due to weak nonlinearity, we assume the timescale separation by introducing the slowly-varying potential $\phi_k=\delta\phi_ke^{i\omega_kt}$ and expanding the linear dielectric constant as $\epsilon_{Ak}\simeq i(\partial\epsilon_{Ak}/\partial\omega_k)\partial_t$ with $\partial\epsilon_{Ak}/\partial\omega_k=2(1-\Gamma_k)/(\omega_kb_k)$. In the light of the wave energy density $E_k=\sigma_k(1-\Gamma_k)|\phi_k|^2$ from Equation (\ref{eqn:generalized_free_energy}), we introduce the wave-action density $n_k=E_k/|\omega_k|$ and  the corresponding wave-action amplitude 
\begin{equation}
    c_k=\sqrt{\frac{\sigma_k(1-\Gamma_k)}{|\omega_k|}}\phi_k\equiv g_k\phi_k. 
\end{equation}
The dynamic equation for wave-action amplitude is then given as 
\begin{equation}\label{eqn:wave-amp-eqn}
    i\partial_tc_k=\int V_{12}^kc_1c_2e^{i\omega_{12}^kt}\delta_{12}^kd\mathbf{k}_1d\mathbf{k}_2,
\end{equation}
where the nonlinear interaction coefficient is defined as
\begin{equation}\label{eqn:nl-interaction-coefficient}
    \begin{aligned}
    V_{12}^k= & \frac{\Lambda_{12}^k}{4|\omega_k|}\frac{\sigma_k\sigma_1\sigma_2}{g_kg_1g_2}\left(\frac{k_{1\parallel}}{\omega_1}-\frac{k_{2\parallel}}{\omega_2}\right)\\ & \times\left(b_k\frac{k_{\parallel}}{\omega_k}+b_1\frac{k_{1\parallel}}{\omega_1}+b_2\frac{k_{2\parallel}}{\omega_2} \right),
    \end{aligned}
\end{equation}
and $\omega_{12}^k\equiv\omega_k-\omega_1-\omega_2$ is the frequency mismatch. 

Since the dispersion of KAWs due to the FILR effect naturally leads to the phase randomization as time progresses on the nonlinear timescale, we apply the ensemble average over random phases on the timescale intermediate between the linear fast oscillation period and the nonlinear timescale, i.e. to adopt the random phase approximation\citep{RZSagdeev1969,VEZakharov1992book}. Assuming all frequencies to be positive, viz. $\omega_k,\omega_1,\omega_2>0$, the kinetic equation of waves in terms of the wave-action density $n_k$\footnote{Contrary to the Eulerian formalism adopted by \cite{TPassotJPP2019} for a gyrofluid model, we apply the canonical Hamiltonian formalism for the present gyrokinetic model using the wave-action density $n_k$ and the wave-action amplitude $c_k$.} is given by 
\begin{equation}\label{eqn:WKE}
    \partial_tn_k=\int d\mathbf{k}_1d\mathbf{k}_2(\mathcal{R}_{12}^k-\mathcal{R}_{2k}^1-\mathcal{R}_{k1}^2),
\end{equation}
with the integrand defined as 
\begin{equation}
    \mathcal{R}_{12}^k=4\pi |V_{12}^k|^2(n_1n_2-n_kn_1-n_kn_2)\delta (\omega_{12}^k)\delta_{12}^k,
\end{equation}
representing the decay process of a plasmon $(\omega_k,\mathbf{k})$ into two plasmons $(\omega_1,\mathbf{k}_1)$, $(\omega_2,\mathbf{k}_2)$, and its reverse merging process. Such a wave kinetic equation (WKE) describes the long-term statistical evolution of an ensemble of random KAWs with broadband spectra. 

Energy and momentum are exact quadratic invariants in the weak-turbulence closure. The existence of the resonant manifold determined by the frequency and wavevector resonant conditions proves the conservation of these quantities. The energy and the parallel-momentum are the two invariants considered in the present weak KAW turbulence of transversally isotropic case, and correspond to the generalized free-energy  and the generalized helicity defined in Equations (\ref{eqn:generalized_free_energy}) and (\ref{eqn:generalized_helicity}), respectively. Similar to the kinetic helicity in three-dimensional hydrodynamic turbulence and the helicities in three-dimensional MHD turbulence\citep{AAlexakisPR2018}, the generalized helicity is also a sign-indefinite quantity and does not constrain the dynamics of the energy cascade in most of the cases. Its mean value equals to zero if the turbulent flow is statistically invariant under mirror symmetry, which corresponds to the balanced case of solar wind turbulence. However, when mirror symmetry is broken, i.e. in the imbalanced case, finite helicity could significantly change the turbulent cascade behaviors. 

In general, there are several classes of triad interactions with different combinations of $(s_k,s_1,s_2)$, where $s_k\equiv sgn(\omega_k/k_{\parallel})=\pm$ denotes the propagation direction if we assume further all parallel wavenumbers to be positive, viz. $k_{\parallel},k_{1\parallel},k_{2\parallel}>0$. 
Thus, the resonance conditions $\omega_k-\omega_1-\omega_2 =0$ and  $s_kk_{\parallel}-s_1k_{1\parallel}-s_2k_{2\parallel}=0$ lead to 
\begin{equation}
    \begin{aligned}
        & \frac{k_{1\parallel}}{k_{\parallel}}=s_ks_1\frac{s_k\alpha_k-s_2\alpha_2}{s_1\alpha_1-s_2\alpha_2},\\ & \frac{k_{2\parallel}}{k_{\parallel}}=s_ks_2\frac{s_k\alpha_k-s_1\alpha_1}{s_2\alpha_2-s_1\alpha_1},
    \end{aligned}
\end{equation}
where the normalized phase velocity $\alpha_k\equiv |\omega_k/(k_{\parallel}v_A)|=\sqrt{\sigma_kb_k/(1-\Gamma_k)}$ is found to be independent of $k_{\parallel}$ and monotonically increase with $k_{\perp}$\citep{LChenRMPP2021,KShenPoP2024}. Therefore, the positiveness of frequencies and parallel wavenumbers leads to the condition of three-wave interaction defined in $\mathcal{R}_{12}^k$: $k_{1\perp}<k_{\perp}<k_{2\perp}$ or $k_{2\perp}<k_{\perp}<k_{1\perp}$ for the co-propagating case $(s_k=s_1=s_2)$, as in Figure \ref{fig:match_cond_co-prop}; $k_{\perp}>k_{1\perp}$ for the counter-propagating case $(s_k=s_1=-s_2)$, as in Figure \ref{fig:match_cond_counter-prop}. Note that the integration regions of $R_{2k}^1$ and $R_{k1}^2$ in the perpendicular wavenumber space are also defined by their resonance conditions as shown in Figures \ref{fig:match_cond_co-prop} and \ref{fig:match_cond_counter-prop}, which coincide with the results obtained from parametric decay analyses\citep{KShenPoP2024}. The boundary lines in these figures are determined by the triangle inequality of perpendicular wavevector matching condition $\mathbf{k}=\mathbf{k}_1+\mathbf{k}_2$. Interestingly, the co-propagating case corresponds to a turbulent flow of which all modes are characterized by the same sign of helicity, such that the total generalized helicity is a sign-definite invariant, i.e. $P=\int k_{\parallel}n_kd\mathbf{k}>0$, in analogy to the enstrophy of two-dimensional hydrodynamic turbulence\citep{GBoffettaARFM2012} and the helicity of three-dimensional helical hydrodynamic turbulence\citep{LBiferalePRL2012,LBiferaleJFM2013,AAlexakisPR2018}. Now that both the energy and the helicity are positive-definite invariants in such a helical KAW turbulence, we expect the system to develop a dual cascade with the energy and the helicity cascading in opposite directions. The possible existence of both an energy cascade and a parallel-momentum/helicity cascade alongwith their stationary power-law solutions in the co-propagating case is a central point of this work.

\subsection{Kolmogorov-Zakharov Spectra}\label{sec:KZ-spectra}

\begin{figure*}[htbp!]
    \centering
    \begin{minipage}{.49\textwidth}
        \centering 
        \includegraphics[width=0.99\linewidth]{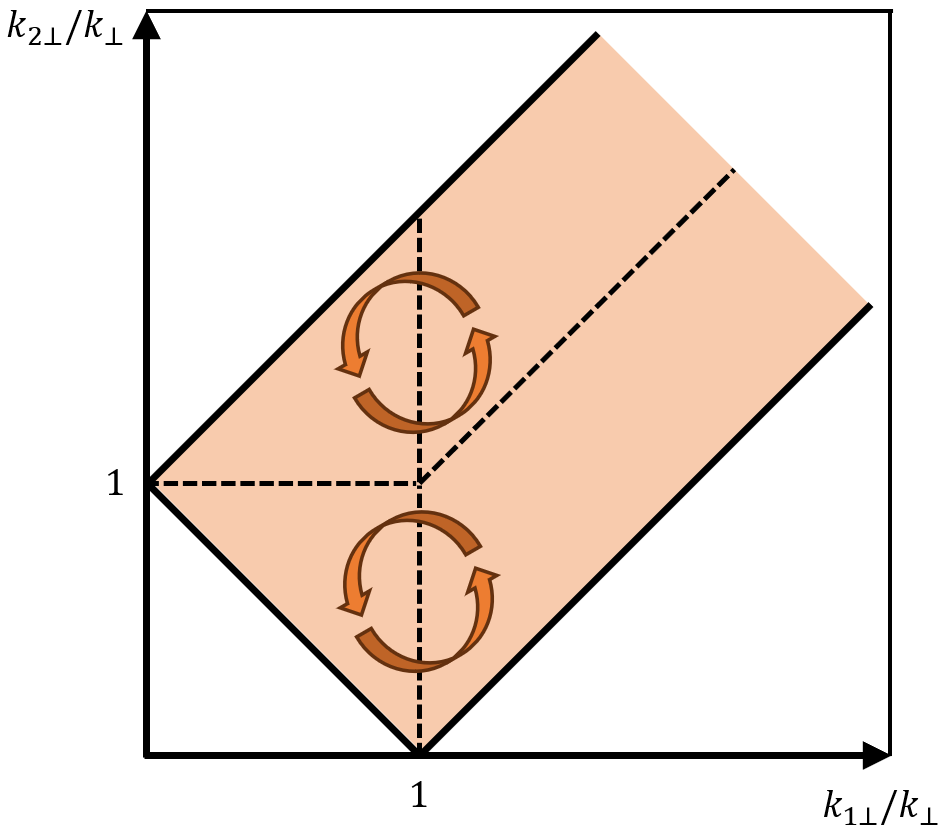}
        \caption{\label{fig:ZT-perp} A schematic illustration of the Zakharov transformation defined in the perpendicular wavenumber space. The colored regions are swapped under the transformation $k_{1\perp}\rightarrow k_{\perp}^2/k_{1\perp}$, $k_{2\perp}\rightarrow k_{\perp}k_{2\perp}/k_{1\perp}$.}
    \end{minipage}
    \begin{minipage}{.49\textwidth}
        \centering 
        \includegraphics[width=0.99\linewidth]{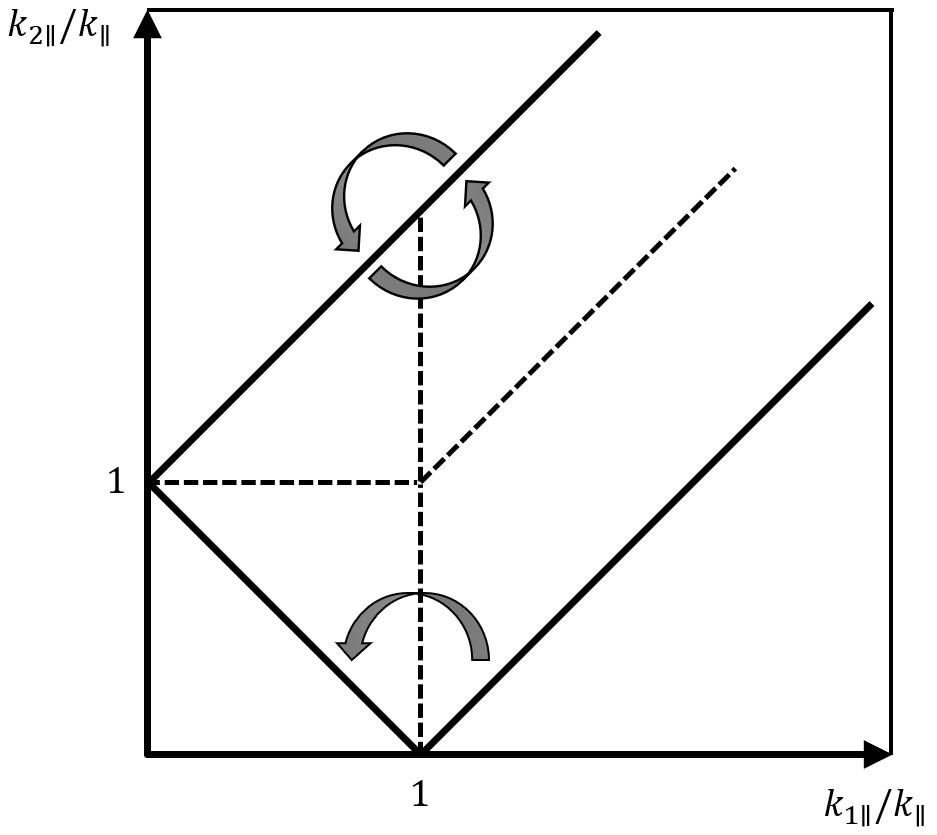}
        \caption{\label{fig:ZT-parallel} A schematic illustration of the Zakharov transformation defined in the parallel wavenumber space. The solid lines are swapped under the transformation $k_{1\parallel}\rightarrow k_{\parallel}^2/k_{1\parallel}$, $k_{2\parallel}\rightarrow k_{\parallel}k_{2\parallel}/k_{1\parallel}$.}
    \end{minipage}
\end{figure*}

To calculate the stationary spectra of a cylindrically symmetric anisotropic medium, one assumes that the medium is scale-invariant (e.g. \cite{VEZakharovJAMTP1967,ANPushkarevPRL1996}) or nearly scale-invariant (e.g. \cite{EAKuznetsovJETP1972,AMBalkPLA1990}) to derive power-law solutions in terms of the magnitudes of wavevectors. Here, anisotropy is introduced in the present model due to the presence of a mean magnetic field, so the dispersion law and the nonlinear interaction coefficient depend differently on the perpendicular and parallel wavenumbers \citep{LChenPoP2013}. The notion of cylindrical symmetry means that the system is transversally isotropic with respect to the mean magnetic field and depends on $k_{\parallel}$ and $|\mathbf{k}_{\perp}|$ only. Scale-invariance means that the statistical properties of nonlinear interactions are universal in the inertial range. Consequently, in the following analysis we assume the frequency and the interaction coefficient to be bi-homogeneous functions denoted as $\omega_k(h_{\parallel}k_{\parallel},h_{\perp}k_{\perp})=h_{\parallel}^{\alpha_{\parallel}}h_{\perp}^{\alpha_{\perp}}\omega_k(k_{\parallel},k_{\perp})$ and $V_{(h_{\parallel}k_{1\parallel},h_{\perp}k_{1\perp})(h_{\parallel}k_{2\parallel},h_{\perp}k_{2\perp})}^{(h_{\parallel}k_{\parallel},h_{\perp}k_{\perp})}=h_{\parallel}^{\beta_{\parallel}}h_{\perp}^{\beta_{\perp}}V_{(k_{1\parallel},k_{1\perp})(k_{2\parallel},k_{2\perp})}^{(k_{\parallel},k_{\perp})}$, where $h_{\parallel}$ and $h_{\perp}$ are positive constants, $(\alpha_{\parallel},\alpha_{\perp})$ are the power indexes of the real frequency, and $(\beta_{\parallel},\beta_{\perp})$ are the homogeneity degrees of $V_{12}^k$ with respect to $(k_{\parallel},k_{\perp})$.
The stationary power-law solutions take the form of $n_k\propto k_{\parallel}^{\nu_{\parallel}}k_{\perp}^{\nu_{\perp}}$ and the exponents $(\nu_{\parallel},\nu_{\perp})$ can be obtained by the Zakharov transformation\citep{VEZakharovJAMTP1967,VEZakharov1992book}. The transformation consists of changes of variables that conformally swap the integration regions in both the perpendicular wavenumber space and the parallel wavenumber space, as shown in Figures \ref{fig:ZT-perp} and \ref{fig:ZT-parallel}. The Zakharov transformation applied to the second integral term of the WKE is given as:
\begin{equation}
    \begin{aligned}
        & k_{1\perp}\rightarrow k_{\perp}^2/k_{1\perp},\quad k_{2\perp}\rightarrow k_{\perp}k_{2\perp}/k_{1\perp},\\
        & k_{1\parallel}\rightarrow k_{\parallel}^2/k_{1\parallel},\quad k_{2\parallel}\rightarrow k_{\parallel}k_{2\parallel}/k_{1\parallel}.
    \end{aligned}
\end{equation}
Based on scaling properties of $\omega_k$, $V_{12}^k$ and $n_k$, the second integral is mapped onto the first integral via the transformation,
\begin{equation}
    \int d \mathbf{k}_1d\mathbf{k}_2\mathcal{R}_{2k}^1\rightarrow \int  d \mathbf{k}_1d\mathbf{k}_2\mathcal{R}_{12}^k\left(\frac{k_{1\parallel}}{k_\parallel}\right)^{\mu_{\parallel}}\left(\frac{k_{1\perp}}{k_{\perp}}\right)^{\mu_{\perp}},
\end{equation}
with the indexes given as:
\begin{equation}
    \begin{aligned}
    & \mu_{\parallel}=\alpha_{\parallel}-2\nu_{\parallel}-2\beta_{\parallel}-2,\\ & \mu_{\perp}=\alpha_{\perp}-2\nu_{\perp}-2\beta_{\perp}-4.
    \end{aligned}
\end{equation}
Similarly the Zakharov transformation is applied onto the third integral, giving an additional exponential factor. Finally we obtain the factorized form of the WKE:
\begin{equation}\label{eqn:WKE-after-ZT}
    \begin{aligned}
    \partial_tn_k= \int d \mathbf{k}_1d\mathbf{k}_2 \mathcal{R}_{12}^k\left[1-\left(\frac{k_{1\parallel}}{k_{\parallel}}\right)^{\mu_{\parallel}}\left(\frac{k_{1\perp}}{k_{\perp}}\right)^{\mu_{\perp}}\right.\\ \left. -\left(\frac{k_{2\parallel}}{k_{\parallel}}\right)^{\mu_{\parallel}}\left(\frac{k_{2\perp}}{k_{\perp}}\right)^{\mu_{\perp}}\right].
    \end{aligned}
\end{equation}

The collision integral on the right-hand-side of the WKE vanishes when the factor in the bracket coincides with the delta functions in $\mathcal{R}_{12}^k$, i.e. $\delta (\omega_k-\omega_1-\omega_2)$ or $\delta (k_{\parallel}-k_{1\parallel}-k_{2\parallel})$, as more clearly indicated by the dimensionless form of collision integral (\ref{eqn:dimensionless-collision-integral}). The resulting stationary power-law spectra, called the Kolmogorov-Zakharov (KZ) spectra, are obtained by equating $(\mu_{\parallel},\mu_{\perp})$ to $(\alpha_{\parallel},\alpha_{\perp})$ or $(1,0)$, corresponding to a constant-flux energy cascade or a constant-flux parallel-momentum cascade, respectively. These choices lead to, respectively,
\begin{equation}\label{eqn:KZ-energy}
    n_k\propto k_{\parallel}^{-\beta_{\parallel}-1}k_{\perp}^{-\beta_{\perp}-2},
\end{equation}
and
\begin{equation}\label{eqn:KZ-momentum}
    n_k\propto k_{\parallel}^{(\alpha_{\parallel}-2\beta_{\parallel}-3)/2}k_{\perp}^{(\alpha_{\perp}-2\beta_{\perp}-4)/2}.
\end{equation}
Note that the parallel-momentum cascade is allowed only in the case of all KAWs are co-propagating, such that the parallel-momentum is sign-definite and maximized at all $\mathbf{k}$. Equations \ref{eqn:KZ-energy} and \ref{eqn:KZ-momentum} are the non-equilibrium solutions carrying finite fluxes in the inertial range, which shall be applied to derive stationary spectra of the KAW turbulence in Section \ref{sec:results}. There is another kind of fluxless solutions corresponding to the thermodynamic equilibrium distribution, which is not of our interest here. 

\subsection{Cascade Direction}\label{sec:Cascade-Direction}

The fluxes and cascade directions of KAW KZ spectra are studied here. 
Due to conservation laws in the inertial range, we write out the continuity equation for the energy density $E(\mathbf{k})=\omega_kn_k$,
\begin{equation}\label{eqn:E-continuity}
    \partial_tE(\mathbf{k})=-\frac{\partial\Pi_{\parallel}}{\partial k_{\parallel}}-\frac{1}{k_{\perp}}\frac{\partial (k_{\perp}\Pi_{\perp})}{\partial k_{\perp}},
\end{equation}
and that for the parallel-momentum density $P(\mathbf{k})=k_{\parallel}n_k$,
\begin{equation}\label{eqn:P-continuity}
    \partial_tP(\mathbf{k})=-\frac{\partial R_{\parallel}}{\partial k_{\parallel}}-\frac{1}{k_{\perp}}\frac{\partial (k_{\perp}R_{\perp})}{\partial k_{\perp}},
\end{equation}
for the case of cylindrical symmetry, where $(\Pi_{\parallel},\Pi_{\perp})$ and $(R_{\parallel},R_{\perp})$ are the parallel and perpendicular fluxes of $E(\mathbf{k})$ and $P(\mathbf{k})$, respectively. We then introduce the dimensionless variables 
$\xi_{1\parallel}\equiv k_{1\parallel}/k_{\parallel}$, $\xi_{1\perp}\equiv k_{1\perp}/k_{\perp}$, $\xi_{2\parallel}\equiv k_{2\parallel}/k_{\parallel}$, $\xi_{2\perp}\equiv k_{2\perp}/k_{\perp}$, and rewrite Equation \ref{eqn:WKE-after-ZT} into 
\begin{equation}
    \partial_tn_k=\frac{4\pi A^2C_V^2}{C_{\omega}}k_{\parallel}^{-\mu_{\parallel}-1}k_{\perp}^{-\mu_{\perp}-2}I_{\xi}(\nu_{\parallel},\nu_{\perp}),
\end{equation}
where $n_k\equiv Ak_{\parallel}^{\nu_{\parallel}}k_{\perp}^{\nu_{\perp}}$, $\omega_k\equiv C_{\omega}k_{\parallel}^{\alpha_{\parallel}}k_{\perp}^{\alpha_{\perp}}$, $|V_{12}^k|=C_Vk_{\parallel}^{\beta_{\parallel}}k_{\perp}^{\beta_{\perp}}|V_{\xi_1\xi_2}|$, and the dimensionless integral is defined as 
\begin{equation}\label{eqn:dimensionless-collision-integral}
    \begin{aligned}
        & I_{\xi}(\nu_{\parallel},\nu_{\perp})= \int\frac{|V_{\xi_1\xi_2}|^2}{\sin\theta_k}(\xi_{1\parallel}\xi_{2\parallel})^{\nu_{\parallel}}(\xi_{1\perp}\xi_{2\perp})^{\nu_{\perp}}\\ & \times (1-\xi_{1\parallel}^{-\nu_{\parallel}}\xi_{1\perp}^{-\nu_{\perp}}-\xi_{2\parallel}^{-\nu_{\parallel}}\xi_{2\perp}^{-\nu_{\perp}})(1-\xi_{1\parallel}^{\mu_{\parallel}}\xi_{1\perp}^{\mu_{\perp}}\\ & -\xi_{2\parallel}^{\mu_{\parallel}}\xi_{2\perp}^{\mu_{\perp}})\delta (1-\xi_{1\parallel}^{\alpha_{\parallel}}\xi_{1\perp}^{\alpha_{\perp}}-\xi_{2\parallel}^{\alpha_{\parallel}}\xi_{2\perp}^{\alpha_{\perp}})\\ & \times\delta (s_k-s_1\xi_{1\parallel}-s_2\xi_{2\parallel})d\xi_{1\parallel}d\xi_{2\parallel}d\xi_{1\perp}d\xi_{2\perp}.
    \end{aligned}
\end{equation}
Note that $\int d\mathbf{k}_1d\mathbf{k}_2\delta_{12}^k=\int d\xi_{1\parallel}d\xi_{1\perp}d\xi_{2\parallel}d\xi_{2\perp}\delta (s_k-s_1\xi_{1\parallel}-s_2\xi_{2\parallel})(k_{\parallel}k_{\perp}^2)/\sin\theta_k$ is used above and the dimensionless variable $|V_{\xi_1\xi_2}|$ can be straightforwardly obtained from Equation \ref{eqn:nl-interaction-coefficient}. Fixing $k_{\parallel}$, integrating the continuity equations with respect to $k_{\perp}$ and making use of the L'Hopital's rule gives the perpendicular fluxes of energy and parallel-momentum:
\begin{equation}
    \Pi_{\perp}=-\frac{2\pi A^2C_V^2}{k_{\parallel}k_{\perp}}\frac{\partial I_{\xi}(\nu_{\parallel},\nu_{\perp})}{\partial\nu_{\perp}}\bigg|_{\substack{\nu_{\parallel}=-\beta_{\parallel}-1\\ \nu_{\perp}=-\beta_{\perp}-2}},
\end{equation}
\begin{equation}
    R_{\perp}=-\frac{2\pi A^2C_V^2}{C_{\omega}k_{\parallel}k_{\perp}}\frac{\partial I_{\xi}(\nu_{\parallel},\nu_{\perp})}{\partial\nu_{\perp}}\bigg|_{\substack{\nu_{\parallel}=(\alpha_{\parallel}-2\beta_{\parallel}-3)/2\\ \nu_{\perp}=(\alpha_{\perp}-2\beta_{\perp}-4)/2}}.
\end{equation}
At constant $k_{\perp}$, we also obtain, after integrating over $k_{\parallel}$:
\begin{equation}
    \Pi_{\parallel}=-\frac{2\pi A^2C_V^2}{k_{\perp}^2}\frac{\partial I_{\xi}(\nu_{\parallel},\nu_{\perp})}{\partial\nu_{\parallel}}\bigg|_{\substack{\nu_{\parallel}=-\beta_{\parallel}-1\\ \nu_{\perp}=-\beta_{\perp}-2}},
\end{equation}
\begin{equation}
    R_{\parallel}=-\frac{2\pi A^2C_V^2}{C_{\omega}k_{\perp}^2}\frac{\partial I_{\xi}(\nu_{\parallel},\nu_{\perp})}{\partial\nu_{\parallel}}\bigg|_{\substack{\nu_{\parallel}=(\alpha_{\parallel}-2\beta_{\parallel}-3)/2\\ \nu_{\perp}=(\alpha_{\perp}-2\beta_{\perp}-4)/2}}.
\end{equation}
The fluxes are predominantly transverse to the plane\citep{VEZakharov1992book,SGaltierJPP2015,VDavidJPP2022}, i.e. $\Pi_{\perp}\gg\Pi_{\parallel}$ and $R_{\perp}\gg R_{\parallel}$, due to the anisotropic assumption $k_{\parallel}\ll k_{\perp}$. Therefore we concentrate on the cascade directions of energy and parallel-momentum in the perpendicular wavenumber space. The expressions and signs of $\Pi_{\perp}$ and $R_{\perp}$ are specified in each case shown below and in Appendix \ref{sec:appendix-cascade-direction}.

\section{Results}\label{sec:results}

\begin{table*}[htbp!]
\begin{ruledtabular}\setlength{\tabcolsep}{15pt}\caption{\label{tab:Summary} The power indexes of the real frequency $(\alpha_{\parallel},\alpha_{\perp})$, the homogeneity degrees of the nonlinear interaction coefficient $(\beta_{\parallel},\beta_{\perp})$, the KZ solutions $(\nu_{\parallel},\nu_{\perp})$ corresponding to energy cascade and parallel-momentum cascade and their cascade directions, in the three regimes, i.e., $k_{\perp}\rho_i\gg 1$, $k_{\perp}\rho_i\ll 1$ (co-propagating), and $k_{\perp}\rho_i\ll 1$ (counter-propagating).}
\begin{tabular}{cccccc}
regime & $(\alpha_{\parallel},\alpha_{\perp})$ & $(\beta_{\parallel},\beta_{\perp})$ & invariant & $(\nu_{\parallel},\nu_{\perp})$ & direction \\
\hline
\multirow{2}{*}{$k_{\perp}\rho_i\gg 1$} & \multirow{2}{*}{(1,1)} & \multirow{2}{*}{(1/2,5/2)} & energy & (-3/2,-9/2) & direct \\ 
 & & & momentum & (-3/2,-4) & inverse \\
\multirow{2}{*}{$k_{\perp}\rho_i\ll 1$ (co-prop)} & \multirow{2}{*}{(1,2)} & \multirow{2}{*}{(1/2,3)} & energy &  (-3/2,-5) & direct \\ 
 & & & momentum & (-3/2,-4) & inverse \\
 $k_{\perp}\rho_i\ll 1$ (counter-prop) & (1,0) & (1/2,1) & energy & (-3/2,-3) & direct \\ 
\end{tabular}
\end{ruledtabular}
\end{table*}
\begin{figure*}[htbp!]
    \centering
    \begin{minipage}{.49\textwidth}
        \centering 
        \includegraphics[width=0.99\linewidth]{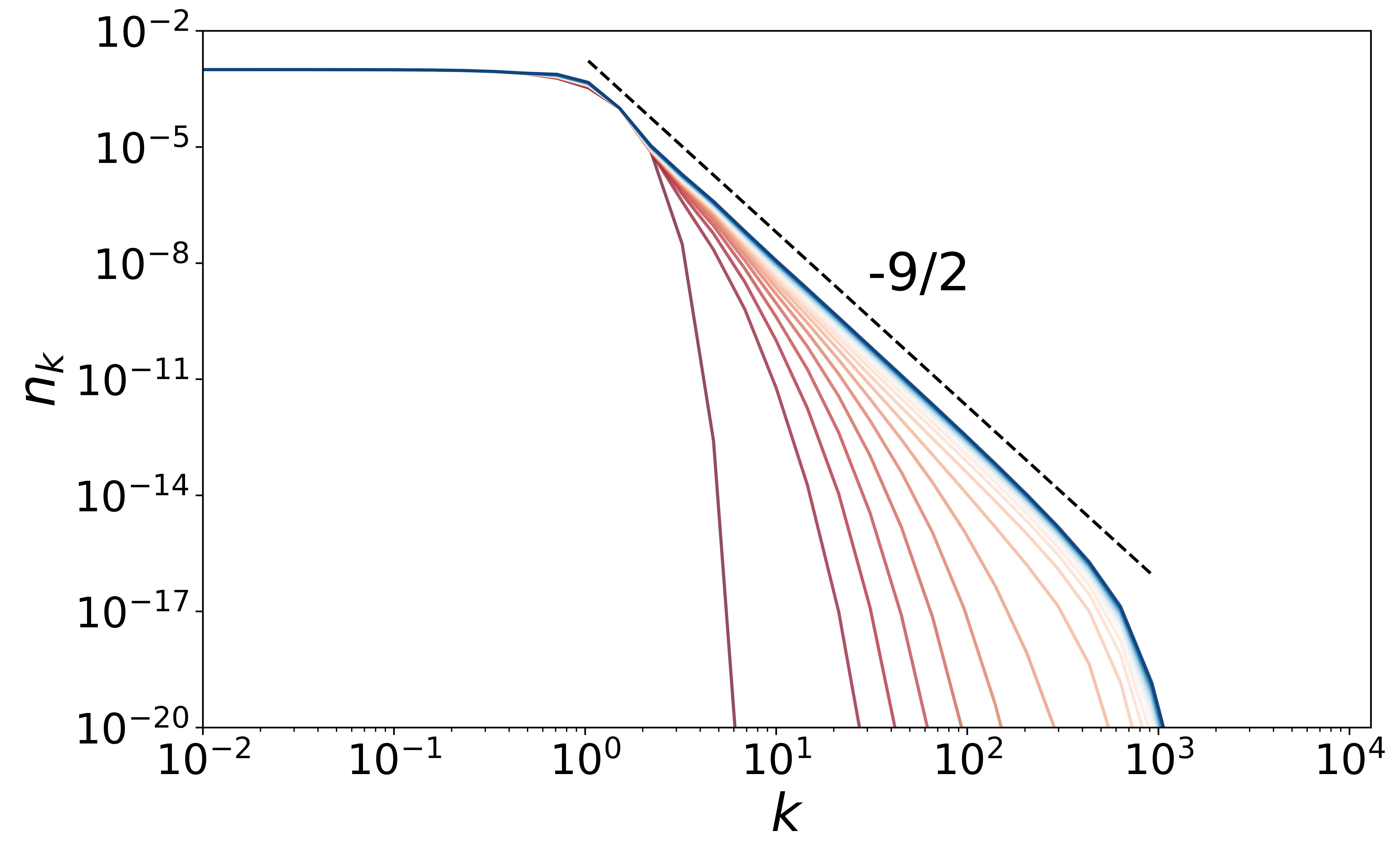}
        \caption{\label{fig:SWL-cascade} The formation of $k_{\perp}^{-9/2}$ stationary spectrum in the short-wavelength limit due to direct energy cascade.}
    \end{minipage}
    \begin{minipage}{.49\textwidth}
        \centering 
        \includegraphics[width=0.99\linewidth]{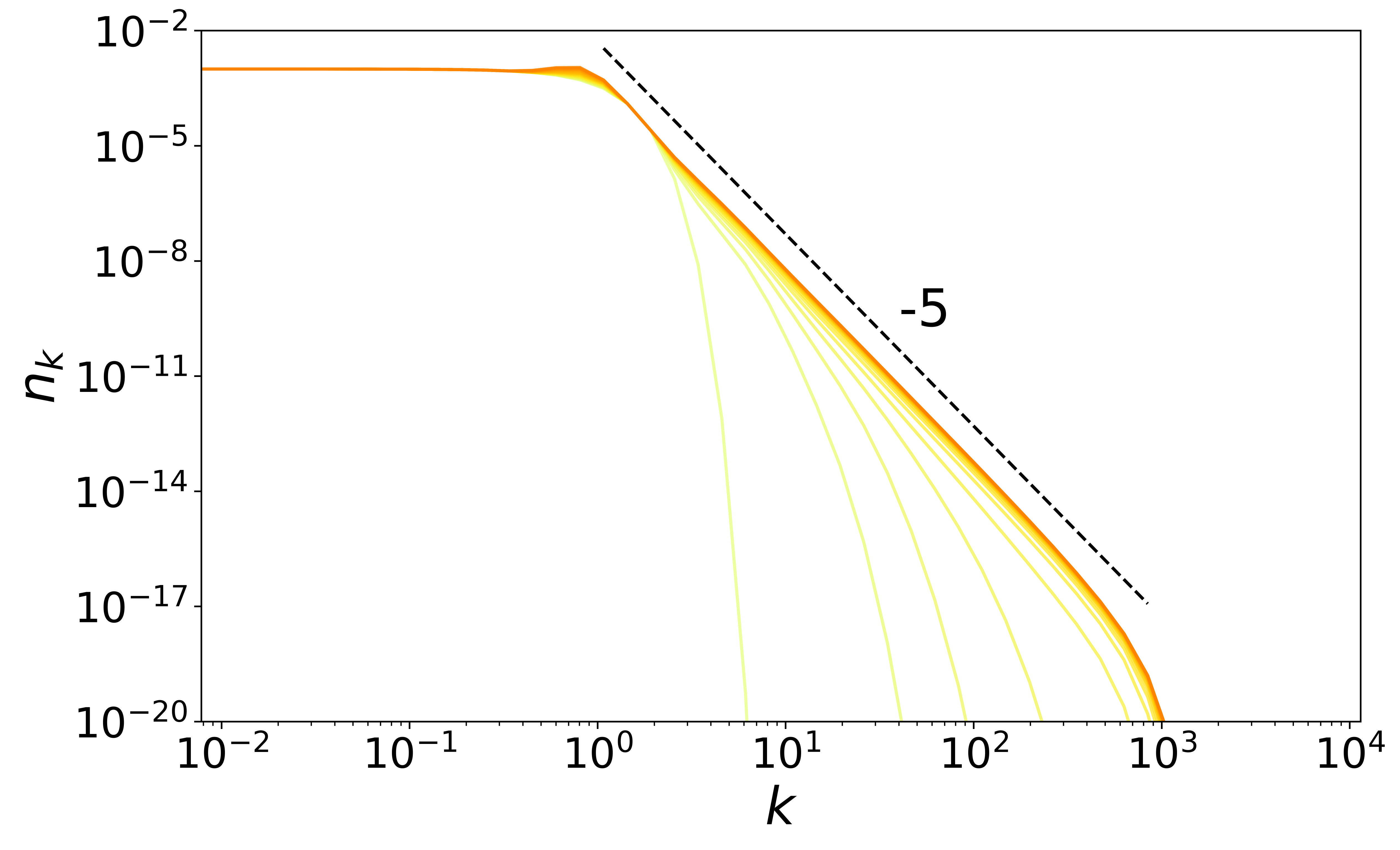}
        \caption{\label{fig:LWL-cascade} The formation of $k_{\perp}^{-5}$ stationary spectrum in the long-wavelength limit due to direct energy cascade.}
    \end{minipage}
\end{figure*}
\begin{figure*}[htbp!]
    \centering
    \begin{minipage}{.49\textwidth}
        \centering 
        \includegraphics[width=0.99\linewidth]{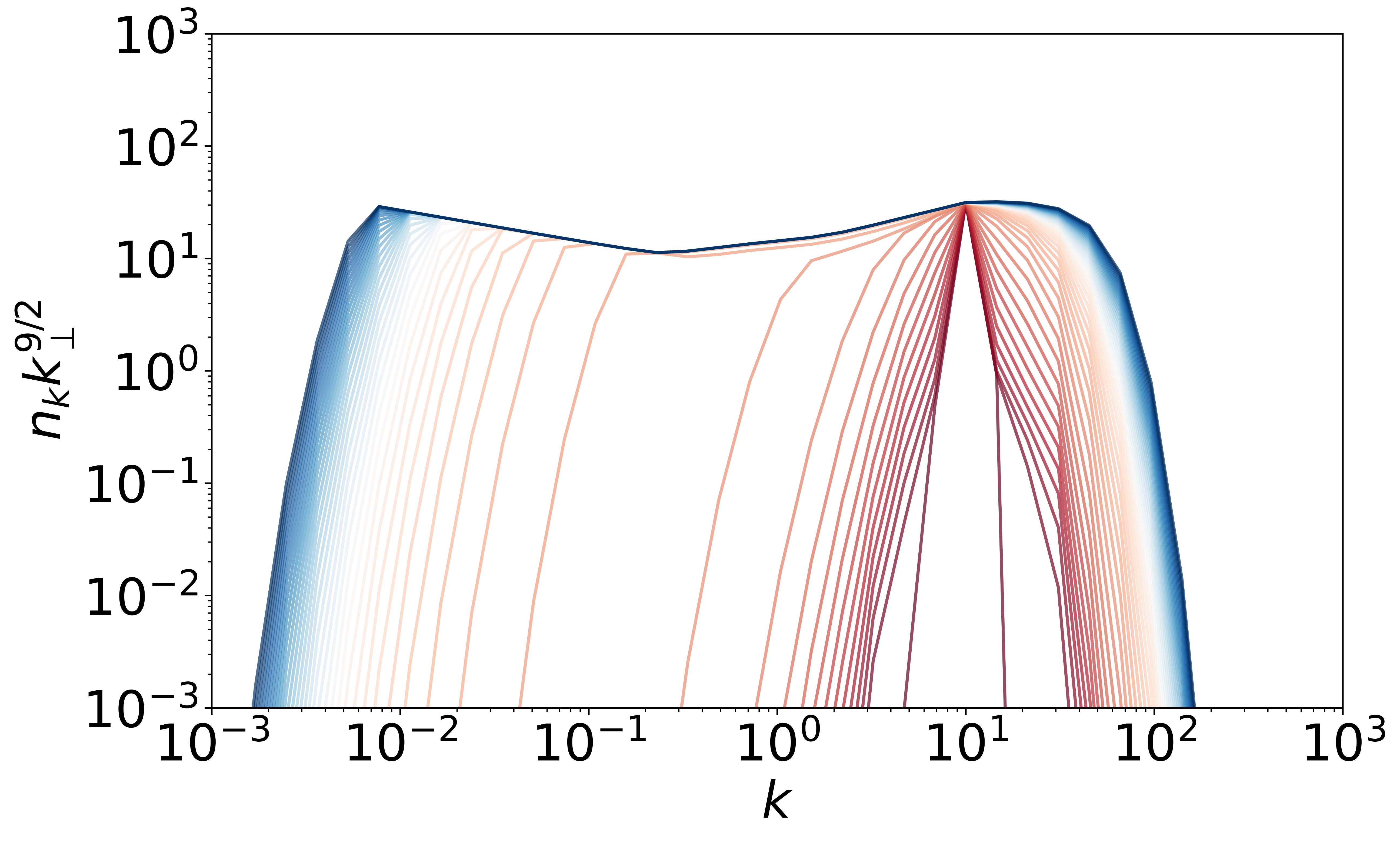}
        \caption{\label{fig:full-FILR-cascade} Temporal evolution of $n(k_{\perp})$ retaining the full FILR effect.}
    \end{minipage}
    \begin{minipage}{.49\textwidth}
        \centering 
        \includegraphics[width=0.99\linewidth]{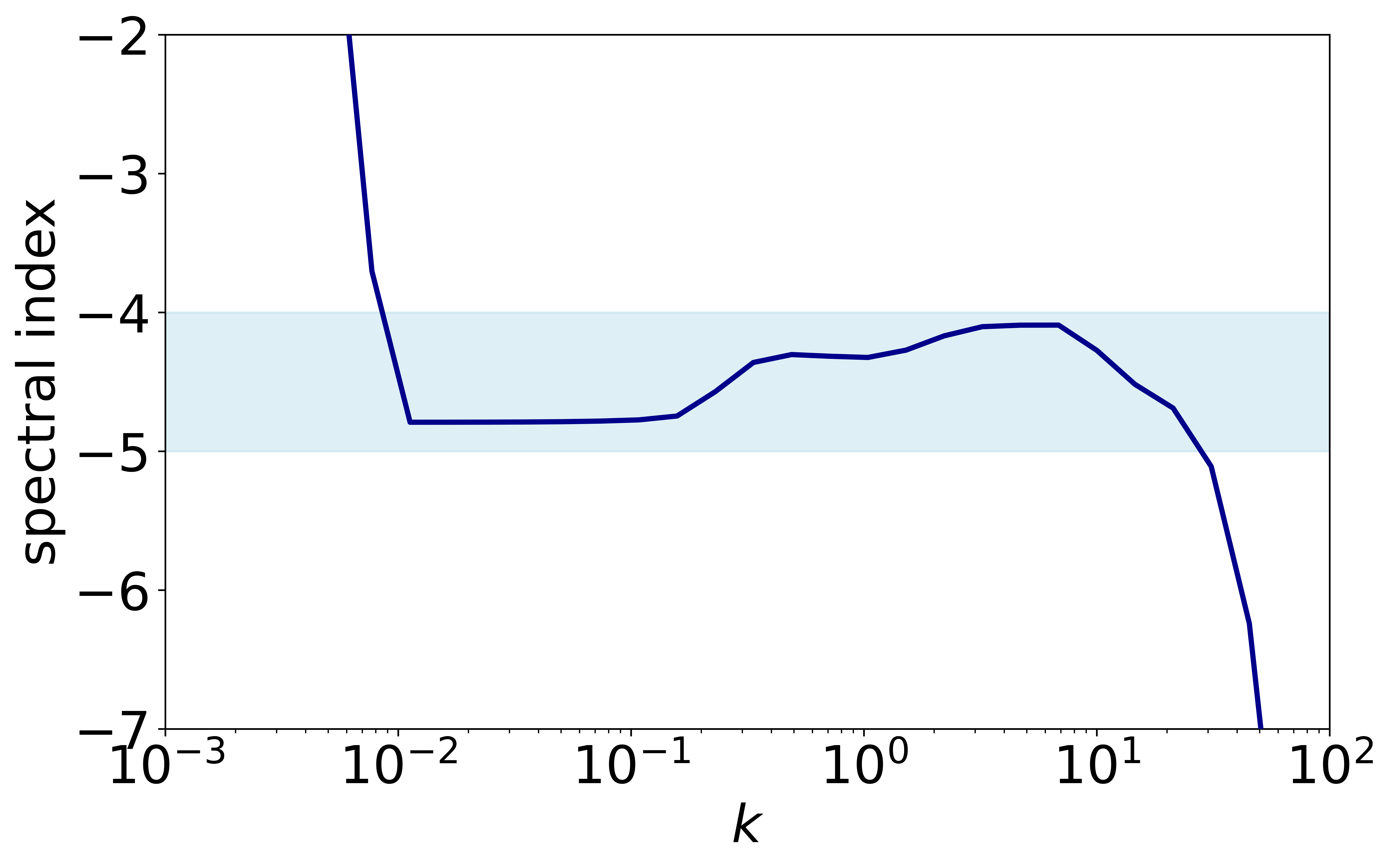}
        \caption{\label{fig:spectral-index-full-FULR cascade} The spectral index of the saturated spectrum obtained in Figure \ref{fig:full-FILR-cascade}.}
    \end{minipage}
\end{figure*}

\subsection{Stationary KAW Spectra and Cascade Direction}

To obtain the KZ spectra of KAW turbulence, we firstly separate the model into two limits, i.e. the short-wavelength limit $(k_{\perp}\rho_i\gg 1)$ and the long-wavelength limit $(k_{\perp}\rho_i\ll 1)$, since the scale-invariance of KAW turbulence is broken by the ion gyroscale which sets a preferred scale in $k_{\perp}$ space. In the following, the WKE shall be analyzed in different parameter regimes for the KAW stationary spectra, noting the different power law dependence of the KAW frequency and nonlinear  interaction coefficients.

\paragraph{Short-wavelength limit}
In the short-wavelength limit which corresponds to the ``{ion-kinetic}'' range of solar wind turbulence, we obtain the scale-invariant real frequency 
\begin{equation}
    \omega_k\propto k_{\parallel}k_{\perp},
\end{equation}
and the interaction coefficient 
\begin{equation}\label{eqn:nl-interaction-coefficient-SWL}
    \begin{aligned}
        |V_{12}^k|\propto & \sin\theta_k\left(\frac{k_{1\parallel}k_{1\perp}k_{2\parallel}k_{2\perp}}{k_{\parallel}k_{\perp}}\right)^{1/2}|s_1k_{2\perp}-s_2k_{1\perp}|\\ & \times |s_kk_{\perp}+s_1k_{1\perp}+s_2k_{2\perp}|,
    \end{aligned}
\end{equation}
by taking $\Gamma_k\simeq 0$, $\sigma_k\simeq 1+\tau$, and $\alpha_k\simeq \sqrt{(1+\tau)b_k}$. All numerical coefficients are omitted for simplicity, with emphasis on scaling dependences. The model equation in the short-wavelength limit coincides with the low-$\beta$ limit of anisotropic electron reduced magnetohydrodynamics (ERMHD)\citep{AASchekochihinApJS2009,SGaltierPoP2003,SGaltierJPP2015}.

\paragraph{Long-wavelength limit} The long-wavelength limit of KAW turbulence is further divided into the co-propagating case and the counter-propagating case, as the nonlinear interaction coefficients are significantly different in these two cases\citep{KShenPoP2024}. For the co-propagating case, we have the nearly scale-invariant real frequency 
\begin{equation}
    \omega_k-k_{\parallel}v_A\propto k_{\parallel}k_{\perp}^2,
\end{equation}
and the interaction coefficient retaining the lowest-order contribution from the FILR effect
\begin{equation}
    \begin{aligned}
        |V_{12}^k|\propto & \frac{\sin\theta_k}{k_{\perp}}\left(\frac{k_{1\parallel}k_{2\parallel}}{k_{\parallel}}\right)^{1/2}|k_{1\perp}^2-k_{2\perp}^2|\\ & \times (k_{\perp}^2+k_{1\perp}^2+k_{2\perp}^2),
    \end{aligned}
\end{equation}
where $\sigma_k\simeq 1+\tau b_k$ and $\alpha_k\simeq 1+(3/4+\tau)b_k$ are substituted. Here, the frequency matching condition reduces to $\delta (k_{\parallel}k_{\perp}^2-k_{1\parallel}k_{1\perp}^2-k_{2\parallel}k_{2\perp}^2)$ where the parallel wavenumber matching condition ensures the leading $k_{\parallel}$ terms to vanish. Similar properties exist in the weakly dispersive ion-acoustic turbulence\citep{EAKuznetsovJETP1972,EAKochurinPRL2024} and the drift/Rossby wave turbulence described by the Charney-Hasegawa-Mima equation\citep{AMBalkPLA1990,AMBalkJETP1990,WHortonRMP1999,CConnaughtonPR2015}. 
However, for the counter-propagating case, the model is simplified into the reduced MHD (RMHD) model describing counter-propagating SAW wave-packets interacting nonlinearly with each other\citep{RHKraichnanPoF1965,CSNgApJ1996,SGaltierApJ2002}. The resulting frequency and the nonlinear interaction coefficient are given as $\omega_k\propto k_{\parallel}$ and $|V_{12}^k|\propto (\sin\theta_k/k_{\perp})(k_{1\parallel}k_{2\parallel}/k_{\parallel})^{1/2}|k_{\perp}^2+k_{1\perp}^2-k_{2\perp}^2|$, respectively. 

The scaling exponents of $\omega_k$ and $V_{12}^k$, and the KZ spectra for the above three cases are summarized in Table \ref{tab:Summary}. 
For the counter-propagating/balanced case, we obtain two KZ spectra corresponding to direct energy-cascade including $n_k\propto k_{\parallel}^{-3/2}k_{\perp}^{-9/2}$ in the $k_{\perp}\rho_i\gg 1$ range and $n_k\propto k_{\parallel}^{-3/2}k_{\perp}^{-3}$ in the $k_{\perp}\rho_i\ll 1$ range. They coincide with the energy spectra $E(k_{\parallel},k_{\perp})\propto k_{\parallel}k_{\perp}^2n_k\propto k_{\parallel}^{-1/2}k_{\perp}^{-5/2}$ derived from ERMHD\citep{SGaltierJPP2015} and $E(k_{\parallel},k_{\perp})\propto k_{\parallel}k_{\perp}n_k\propto k_{\parallel}^{-1/2}k_{\perp}^{-2}$ described by RMHD\citep{SGaltierApJ2002}.
For the case of all KAWs being co-propagating, the KZ solutions corresponding to energy cascade and parallel-momentum cascade are $n_k\propto k_{\parallel}^{-3/2}k_{\perp}^{-9/2}$ and $n_k\propto k_{\parallel}^{-3/2}k_{\perp}^{-4}$ in the $k_{\perp}\rho_i\gg 1$ limit, and transition to $n_k\propto k_{\parallel}^{-3/2}k_{\perp}^{-5}$ and $n_k\propto k_{\parallel}^{-3/2}k_{\perp}^{-4}$ in the $k_{\perp}\rho_i\ll 1$ limit. 
The energy-cascade solution in the $k_{\perp}\rho_i\ll 1$ limit for the co-propagating case is distinct from that for the counter-propagating case due to the differences on the frequency resonance condition and the interaction coefficient\citep{KShenPoP2024}. 
It is found that the conservation of parallel-momentum leads to an inverse cascade, as $\Pi_{\perp}$ and $R_{\perp}$ are calculated in Appendix \ref{sec:appendix-cascade-direction} by substituting in $|V_{\xi_1\xi_2}|$, $(\alpha_{\parallel},\alpha_{\perp})$ and $(\beta_{\parallel},\beta_{\perp})$ for each stationary KZ spectrum. 
Since the direct energy-cascade spectra for the balanced weak KAW turbulence have been verified separately in the long-wavelength limit\citep{SBoldyrevPRL2009,SGaltierApJ2010} and the short-wavelength limit\citep{VDavidPRL2024}, the KZ solutions and the inverse cascade behavior for the co-propagating case, as the new results of this work, are worth further investigation. 

\subsection{Numerical Verification}

For numerical illustration, we consider the evolution of the wave-action density in the perpendicular wavenumber space by assuming $n_k\propto k_{\parallel}^{-3/2}n(k_{\perp})$, since the perpendicular cascade is predominant over the parallel cascade and we suppose the $k_{\parallel}$-dependence of $n_k$ to be close to the stationary regime. In order to verify the predicted stationary spectra, we perform three numerical simulations of the WKE in the co-propagating case $(s_k=s_1=s_2)$, as given in Appendix \ref{sec:appendix-numerics-WKE}, which include the short-wavelength limit, the long-wavelength limit, and the general case where Equation \ref{eqn:nl-interaction-coefficient} is used. An eighth-order hyper-viscous term $\nu k_{\perp}^8$ acting on $n(k_{\perp})$ is supplemented in the equations as a sink for the wave-action density. In our calculations the hyper-viscosity is fixed to $\nu =10^{-22}$. The use of a logarithmic discretization is favored for $k_{\perp}$-grid implementation and we take $k_{\perp}^{(i)}=2^{i/36-13}$ where $i\in [0,960]$. For the time integration, we use the fourth-order Adams-Bashforth numerical scheme. Note that the wave-action density is not a conserved quantity of the WKE, therefore the astringency of the equations is uncertained. The formation of stationary spectra in this case is a preliminary verification of the existence of Kolmogorov-Zakharov spectra.

For the first two simulations, the initial condition is taken as $n(k_{\perp})=10^{-3}\exp (-k_{\perp}^2)$. The temporal evolution of $n(k_{\perp})$ is given in Figures \ref{fig:SWL-cascade} and \ref{fig:LWL-cascade}. We observe that the wave-action spectrum expands towards small scales and reaches the stationary spectra $k_{\perp}^{-9/2}$ and $k_{\perp}^{-5}$ in the short-wavelength limit and the long-wavelength limit respectively, as predicted by the theory. As an example of the general case retaining the complete FILR effect, we initially choose a small-scale injection $n(k_{\perp})=10^{-4}\exp [-(k_{\perp}-10)^2/4]$ and take $\rho_i=1$ without loss of generality. The result is shown in Figure \ref{fig:full-FILR-cascade}. We see that the compensated spectra exhibit a dual cascade with the wave-action density transferring in opposite directions and the spectral slope is plotted in Figure \ref{fig:spectral-index-full-FULR cascade}. Below the scale of injection, a narrow spectrum with power index close to $-9/2$ is obtained. However, at large scales we observe that an inverse transfer of $n(k_{\perp})$ leads to a stationary spectrum with spectral index varying approximately from $-4$ to $-5$, which falls within the theoretical prediction. Besides, the formation of a steeper spectrum at larger scales implies the potential existence of nonlocal transfer behaviors. The locality of nonlinear interactions and cross-scale cascade behaviors of the weak KAW turbulence shall be investigated in future works. 

\section{Conclusion and Discussion}\label{sec:conclusion-discussion}

We have proposed the wave-kinetic description of weak KAW turbulence based on the nonlinear mode equation derived from the gyrokinetic theory. The wave kinetic equation (WKE) describing KAW spectral evolution due to resonant three-wave interaction is derived, which is then solved analytically for stationary spectra following the  Zakharov-transformation approach,  and the obtained results are verified by direct numerical solution of the WKE.

The results of stationary power-law spectra are summarized as: (i) using Zakharov transformations and Kolmogorov-Zakharov spectra for cylindrically symmetric systems, we predict the possible stationary power-law solutions in the short and long-wavelength limit for both counter-propagating and co-propagating cases as shown in Table \ref{tab:Summary}; (ii) the direct energy-cascade spectra for the co-propagating case, i.e. $n(k_{\perp})\propto k_{\perp}^{-5}$ and $n(k_{\perp})\propto k_{\perp}^{-9/2}$ in the long and short-wavelength limit respectively, are verified by numerically calculating the WKE; (iii) the direct energy-cascade spectra for co-propagating KAWs in the long-wavelength limit gives $E(k_{\perp})\propto k_{\perp}n(k_{\perp})\propto k_{\perp}^{-4}$ and serves as a potential explanation for the steeper spectrum in the transition range of the solar wind turbulence due to the dispersive nature of KAWs and the enhanced nonlinearity\citep{TABowenPRL2020,SYHuangApJL2021,TPassotJPP2022,TPassotFPP2024}; (iv) the inverse cascade for the co-propagating case is identified numerically with a small-scale injection, although the spectral index does not coincide well with the theory; (v) the later stage of inverse transfer at large scales exhibits a seemingly nonlocal signature\citep{GMiloshevichJPP2021}.

Although we concentrate on the quasi-particle picture of KAW turbulence in this work, the inverse cascade is expected in both short and long-wavelength limit of weak and strong KAW turbulence. The analytical solutions in the weak regime ought to be produced in simulations to prove that they are truly attractive. Whether the conservation of parallel-momentum leads to local cascade spectrum or nonlocal energy transfer is left for future numerical simulations of the WKE and/or the gyrokinetic model. On the latter point, a helical Fourier decomposition\citep{LBiferalePRL2012,LBiferaleJFM2013} focusing on the nonlinear interactions among co-propagating KAWs, which is a natural subset of the nonlinear evolution of KAW turbulence, could be feasible. Regarding the applicable region of the weak-nonlinearity assumption for the weak turbulence closure, one needs to verify the formation of the direct energy-cascade spectra of weak KAW turbulence across the ion gyroscale and the possibility of the coexistence of weak and strong turbulence.

The present work, to suppress the compressional Alfv\'en wave related physics, is formulated in the $1\gg \beta_e\sim\beta_i\gg m_e/m_i$ limit, and a gyrokinetic theory based weak-turbulence approach is adopted. These assumptions may limit the application of the present theory. Nonetheless, the low-$\beta$ assumption still holds well in numerous space and astrophysical plasmas, such as the near-Sun solar wind and the solar corona\citep{RBrunoLRSP2013}. Generalization of the present analysis, including the wave-kinetic approach on KAW turbulence, will be addressed in future publications. 

\begin{acknowledgments}
This work was supported by National Key R\&D Program
of China under Grant No. 2024YFE03170000,
the Strategic Priority Research Program of Chinese
Academy of Sciences under Grant No. XDB0790201,
the National Science Foundation of China under Grant
Nos. 12275236 and 12261131622, and Italian Ministry
for Foreign Affairs and International Cooperation
Project under Grant No. CN23GR02.
\end{acknowledgments}

\appendix

\section{Determination of cascade directions}\label{sec:appendix-cascade-direction}

The cascade direction of each KZ solution is identified here specifically for the co-propagating case. Based on Equation \ref{eqn:dimensionless-collision-integral}, we obtain the full expressions for the perpendicular fluxes:
\begin{equation}
    \begin{aligned}
        \Pi_{\perp}= & -\frac{4\pi A^2C_V^2}{k_{\parallel}k_{\perp}} \int\frac{|V_{\xi_1\xi_2}|^2}{\sin\theta_k}(\xi_{1\parallel}\xi_{2\parallel})^{\nu_{\parallel}}(\xi_{1\perp}\xi_{2\perp})^{\nu_{\perp}}(1-\xi_{1\parallel}^{-\nu_{\parallel}}\xi_{1\perp}^{-\nu_{\perp}}-\xi_{2\parallel}^{-\nu_{\parallel}}\xi_{2\perp}^{-\nu_{\perp}})\\ & \times (\xi_{1\parallel}^{\alpha_{\parallel}}\xi_{1\perp}^{\alpha_{\perp}}ln\xi_{1\perp}+\xi_{2\parallel}^{\alpha_{\parallel}}\xi_{2\perp}^{\alpha_{\perp}}ln\xi_{2\perp})\delta (1-\xi_{1\parallel}^{\alpha_{\parallel}}\xi_{1\perp}^{\alpha_{\perp}}-\xi_{2\parallel}^{\alpha_{\parallel}}\xi_{2\perp}^{\alpha_{\perp}})\\ & \times\delta (1-\xi_{1\parallel}-\xi_{2\parallel})d\xi_{1\parallel}d\xi_{2\parallel}d\xi_{1\perp}d\xi_{2\perp},
    \end{aligned}
\end{equation}
\begin{equation}
    \begin{aligned}
        R_{\perp}= & -\frac{4\pi A^2C_V^2}{C_{\omega}k_{\parallel}k_{\perp}}\int\frac{|V_{\xi_1\xi_2}|^2}{\sin\theta_k}(\xi_{1\parallel}\xi_{2\parallel})^{\nu_{\parallel}}(\xi_{1\perp}\xi_{2\perp})^{\nu_{\perp}}(1-\xi_{1\parallel}^{-\nu_{\parallel}}\xi_{1\perp}^{-\nu_{\perp}}-\xi_{2\parallel}^{-\nu_{\parallel}}\xi_{2\perp}^{-\nu_{\perp}})\\ & \times (\xi_{1\parallel}ln\xi_{1\perp}+\xi_{2\parallel}ln\xi_{2\perp})\delta (1-\xi_{1\parallel}^{\alpha_{\parallel}}\xi_{1\perp}^{\alpha_{\perp}}-\xi_{2\parallel}^{\alpha_{\parallel}}\xi_{2\perp}^{\alpha_{\perp}})\\ & \times\delta (1-\xi_{1\parallel}-\xi_{2\parallel})d\xi_{1\parallel}d\xi_{2\parallel}d\xi_{1\perp}d\xi_{2\perp}.
    \end{aligned}
\end{equation}
We then integrate over $\xi_{1\parallel}$ and $\xi_{2\parallel}$ making use of $\delta (g(x))=\delta (x)/|g(x)'|$. As a result, in the short-wavelength limit we have:
\begin{equation}
    \begin{aligned}
        \Pi_{\perp}= & -\frac{4\pi A^2C_V^2}{k_{\parallel}k_{\perp}} \int \sin\theta_k(\xi_{1\parallel}\xi_{2\parallel}\xi_{1\perp}\xi_{2\perp})|\xi_{2\perp}-\xi_{1\perp}|(1+\xi_{1\perp}+\xi_{2\perp})^2 \\ & \times (\xi_{1\parallel}\xi_{2\parallel})^{-3/2}(\xi_{1\perp}\xi_{2\perp})^{-9/2}(1-\xi_{1\parallel}^{3/2}\xi_{1\perp}^{9/2}-\xi_{2\parallel}^{3/2}\xi_{2\perp}^{9/2})\\ & \times (\xi_{1\parallel}\xi_{1\perp}ln\xi_{1\perp}+\xi_{2\parallel}\xi_{2\perp}ln\xi_{2\perp})d\xi_{1\perp}d\xi_{2\perp},
    \end{aligned}
\end{equation}
\begin{equation}
    \begin{aligned}
        R_{\perp}= & -\frac{4\pi A^2C_V^2}{C_{\omega}k_{\parallel}k_{\perp}}\int\sin\theta_k(\xi_{1\parallel}\xi_{2\parallel}\xi_{1\perp}\xi_{2\perp})|\xi_{2\perp}-\xi_{1\perp}|(1+\xi_{1\perp}+\xi_{2\perp})^2\\ & \times (\xi_{1\parallel}\xi_{2\parallel})^{-3/2}(\xi_{1\perp}\xi_{2\perp})^{-4}(1-\xi_{1\parallel}^{3/2}\xi_{1\perp}^{4}-\xi_{2\parallel}^{3/2}\xi_{2\perp}^{4})\\ & \times (\xi_{1\parallel}ln\xi_{1\perp}+\xi_{2\parallel}ln\xi_{2\perp})d\xi_{1\perp}d\xi_{2\perp},
    \end{aligned}
\end{equation}
where we should insert $\xi_{1\parallel}=(1-\xi_{2\perp})/(\xi_{1\perp}-\xi_{2\perp})$ and $\xi_{2\parallel}=(1-\xi_{1\perp})/(\xi_{2\perp}-\xi_{1\perp})$.
The fluxes in the long-wavelength limit for the co-propagating case are given as:
\begin{equation}
    \begin{aligned}
        \Pi_{\perp}= & -\frac{4\pi A^2C_V^2}{k_{\parallel}k_{\perp}} \int \sin\theta_k\xi_{1\parallel}\xi_{2\parallel}|\xi_{1\perp}^2-\xi_{2\perp}^2|(1+\xi_{1\perp}^2+\xi_{2\perp}^2)^2(\xi_{1\parallel}\xi_{2\parallel})^{-3/2}(\xi_{1\perp}\xi_{2\perp})^{-5}\\ & \times (1-\xi_{1\parallel}^{3/2}\xi_{1\perp}^{5}-\xi_{2\parallel}^{3/2}\xi_{2\perp}^{5})(\xi_{1\parallel}\xi_{1\perp}^{2}ln\xi_{1\perp}+\xi_{2\parallel}\xi_{2\perp}^{2}ln\xi_{2\perp})d\xi_{1\perp}d\xi_{2\perp},
    \end{aligned}
\end{equation}
\begin{equation}
    \begin{aligned}
        R_{\perp}= & -\frac{4\pi A^2C_V^2}{C_{\omega}k_{\parallel}k_{\perp}}\int\sin\theta_k\xi_{1\parallel}\xi_{2\parallel}|\xi_{1\perp}^2-\xi_{2\perp}^2|(1+\xi_{1\perp}^2+\xi_{2\perp}^2)^2(\xi_{1\parallel}\xi_{2\parallel})^{-3/2}(\xi_{1\perp}\xi_{2\perp})^{-4}\\ & \times (1-\xi_{1\parallel}^{3/2}\xi_{1\perp}^{4}-\xi_{2\parallel}^{3/2}\xi_{2\perp}^{4})(\xi_{1\parallel}ln\xi_{1\perp}+\xi_{2\parallel}ln\xi_{2\perp})d\xi_{1\perp}d\xi_{2\perp},
    \end{aligned}
\end{equation}
with $\xi_{1\parallel}=(1-\xi_{2\perp}^2)/(\xi_{1\perp}^2-\xi_{2\perp}^2)$ and $\xi_{2\parallel}=(1-\xi_{1\perp}^2)/(\xi_{2\perp}^2-\xi_{1\perp}^2)$. The above expressions can be easily integrated numerically with respect to $\xi_{1\perp}$ and $\xi_{2\perp}$, to demonstrate that the energy flux is positive and the parallel-momentum flux is negative. 

\section{Numerical calculations of WKE}\label{sec:appendix-numerics-WKE}

The evolution equations of $n(k_{\perp})$ for the co-propagating case are given here. In each case, $\xi_{1\parallel}$ and $\xi_{2\parallel}$ are integrated using the frequency matching condition and the parallel wavenumber matching condition. All the constant numerical coefficients related, e.g. to $\omega_k$, $V_{12}^k$ and $k_{\parallel}$ are omitted for simplicity. 
In the short-wavelength limit, the following equation is numerically calculated:
\begin{equation}\begin{aligned} \partial_tn(k_{\perp})= & \int dk_{1\perp}dk_{2\perp}\sin\theta_k\frac{k_{1\perp}k_{2\perp}}{k_{\perp}}|k_{2\perp}-k_{1\perp}|(k_{\perp}+k_{1\perp}+k_{2\perp})^2\bigg\{[\xi_{1\parallel}\xi_{2\parallel}(\xi_{1\parallel}^{-3/2}\xi_{2\parallel}^{-3/2}n(k_{1\perp})n(k_{2\perp})\\ & -\xi_{1\parallel}^{-3/2}n(k_{\perp})n(k_{1\perp})-\xi_{2\parallel}^{-3/2}n(k_{\perp})n(k_{2\perp}))]|_{\xi_{1\parallel}=\frac{k_{\perp}-k_{2\perp}}{k_{1\perp}-k_{2\perp}}, \xi_{2\parallel}=1-\xi_{1\parallel}}\\ & -2[\xi_{1\parallel}\xi_{2\parallel}(\xi_{2\parallel}^{-3/2}n(k_{2\perp})n(k_{\perp})-\xi_{1\parallel}^{-3/2}\xi_{2\parallel}^{-3/2}n(k_{1\perp})n(k_{2\perp})\\ & -\xi_{1\parallel}^{-3/2}n(k_{1\perp})n(k_{\perp}))]|_{\xi_{1\parallel}=\frac{k_{\perp}-k_{2\perp}}{k_{1\perp}-k_{2\perp}},\xi_{2\parallel}=\xi_{1\parallel}-1}\bigg\},\end{aligned}\end{equation}
where the two terms in the curly brackets correspond to different triad interactions. 
Similarly, the calculated equation in the long-wavelength limit is given as:
\begin{equation}\begin{aligned} \partial_tn(k_{\perp})= & \int dk_{1\perp}dk_{2\perp}\frac{\sin\theta_k}{k_{\perp}^2}|k_{2\perp}^2-k_{1\perp}^2|(k_{\perp}^2+k_{1\perp}^2+k_{2\perp}^2)^2\bigg\{[\xi_{1\parallel}\xi_{2\parallel}(\xi_{1\parallel}^{-3/2}\xi_{2\parallel}^{-3/2}n(k_{1\perp})n(k_{2\perp})\\ & -\xi_{1\parallel}^{-3/2}n(k_{\perp})n(k_{1\perp})-\xi_{2\parallel}^{-3/2}n(k_{\perp})n(k_{2\perp}))]|_{\xi_{1\parallel}=\frac{k_{\perp}^2-k_{2\perp}^2}{k_{1\perp}^2-k_{2\perp}^2},\xi_{2\parallel}=1-\xi_{1\parallel}}\\ & -2[\xi_{1\parallel}\xi_{2\parallel}(\xi_{2\parallel}^{-3/2}n(k_{2\perp})n(k_{\perp})-\xi_{1\parallel}^{-3/2}\xi_{2\parallel}^{-3/2}n(k_{1\perp})n(k_{2\perp})\\ & -\xi_{1\parallel}^{-3/2}n(k_{1\perp})n(k_{\perp}))]|_{\xi_{1\parallel}=\frac{k_{\perp}^2-k_{2\perp}^2}{k_{1\perp}^2-k_{2\perp}^2},\xi_{2\parallel}=\xi_{1\parallel}-1}\bigg\}.\end{aligned}\end{equation}
In the general case where the complete FILR effect is included and $\rho_i=1$ is taken, we simulate the following equation:
\begin{equation}\begin{aligned} \partial_tn(k_{\perp})= & \int dk_{1\perp}dk_{2\perp}\frac{k_{1\perp}^2k_{2\perp}^2\sin\theta_k}{\alpha_k\alpha_1\alpha_2}\frac{\sigma_k\sigma_1\sigma_2}{(1-\Gamma_k)(1-\Gamma_1)(1-\Gamma_2)}|\alpha_2-\alpha_1|\left(\frac{b_k}{\alpha_k}+\frac{b_1}{\alpha_1}+\frac{b_2}{\alpha_2} \right)^2\bigg\{[\xi_{1\parallel}\xi_{2\parallel}(\xi_{1\parallel}^{-3/2}\xi_{2\parallel}^{-3/2}\\ & \times n(k_{1\perp})n(k_{2\perp})-\xi_{1\parallel}^{-3/2}n(k_{\perp})n(k_{1\perp})-\xi_{2\parallel}^{-3/2}n(k_{\perp})n(k_{2\perp}))]|_{\xi_{1\parallel}=\frac{\alpha_k-\alpha_2}{\alpha_1-\alpha_2},\xi_{2\parallel}=1-\xi_{1\parallel}}\\ & -2[\xi_{1\parallel}\xi_{2\parallel}(\xi_{2\parallel}^{-3/2}n(k_{2\perp})n(k_{\perp})-\xi_{1\parallel}^{-3/2}\xi_{2\parallel}^{-3/2}n(k_{1\perp})n(k_{2\perp})\\ & -\xi_{1\parallel}^{-3/2}n(k_{1\perp})n(k_{\perp}))]|_{\xi_{1\parallel}=\frac{\alpha_k-\alpha_2}{\alpha_1-\alpha_2},\xi_{2\parallel}=\xi_{1\parallel}-1}\bigg\}.\end{aligned}\end{equation}


\bibliography{ref}{}
\bibliographystyle{aasjournalv7}



\end{document}